%
\let\useblackboard=\iftrue
%
%
\newfam\black
\input harvmac.tex
\def\Title#1#2{\rightline{#1}
\ifx\answ\bigans\nopagenumbers\pageno0\vskip1in%
\baselineskip 15pt plus 1pt minus 1pt
\else
\def\listrefs{\footatend\vskip 1in\immediate\closeout\rfile\writestoppt
\baselineskip=14pt\centerline{{\bf References}}\bigskip{\frenchspacing%
\parindent=20pt\escapechar=` \input
refs.tmp\vfill\eject}\nonfrenchspacing}
\pageno1\vskip.8in\fi \centerline{\titlefont #2}\vskip .5in}

\ifx\answ\bigans\def\tcbreak#1{}\else\def\tcbreak#1{\cr&{#1}}\fi
\useblackboard
\message{If you do not have msbm (blackboard bold) fonts,}
\message{change the option at the top of the tex file.}
\font\blackboard=msbm10 scaled \magstep1
\font\blackboards=msbm7
\font\blackboardss=msbm5
\textfont\black=\blackboard
\scriptfont\black=\blackboards
\scriptscriptfont\black=\blackboardss

\else

\fi
%
\def\yboxit#1#2{\vbox{\hrule height #1 \hbox{\vrule width #1
\vbox{#2}\vrule width #1 }\hrule height #1 }}
\def\fillbox#1{\hbox to #1{\vbox to #1{\vfil}\hfil}}
\def\ybox{{\lower 1.3pt \yboxit{0.4pt}{\fillbox{8pt}}\hskip-0.2pt}}
\def\np#1#2#3{Nucl. Phys. {\bf B#1} (#2) #3}
\def\pl#1#2#3{Phys. Lett. {\bf #1B} (#2) #3}

\def\physrev#1#2#3{Phys. Rev. {\bf D#1} (#2) #3}

\def\comments#1{}

\def\half{{1\over 2}}
\def\Tr{{{\rm Tr\  }}}

\def\CM{{\cal M}}

\def\CL{{\cal L}}

\def\CW{{\cal W}}

\def\II{\relax{I\kern-.07em I}}

\batchmode
  \font\bbbfont=msbm10
\errorstopmode
\newif\ifamsf\amsftrue
\ifx\bbbfont\nullfont
  \amsffalse
\fi
\ifamsf
\def\IR{\hbox{\bbbfont R}}
\def\IZ{\hbox{\bbbfont Z}}
\def\IF{\hbox{\bbbfont F}}
\def\IP{\hbox{\bbbfont P}}
\else
\def\IR{\relax{\rm I\kern-.18em R}}
\def\IZ{\relax\ifmmode\hbox{Z\kern-.4em Z}\else{Z\kern-.4em Z}\fi}
\def\IF{\relax{\rm I\kern-.18em F}}
\def\IP{\relax{\rm I\kern-.18em P}}
\fi

\def\IR{\relax{\rm I\kern-.18em R}}

\def\BZ{\IZ}
\def\BR{\IR}

\def\bT{{\bf T}}
\def\bS{{\bf S}}

\def\tilde{\widetilde}
\Title{ \vbox{\baselineskip12pt\hbox{hep-th/9705117}
\hbox{RU-97-7}}}
{\vbox{\centerline{Notes on Theories with 16 Supercharges${}^*$}}}

\footnote{}{${}^*$This is an updated version of lectures presented at
the Jerusalem Winter School on Strings and Duality.  These notes will
appear in the Proceedings of The Trieste Spring School (March 7--12).} 

\centerline{Nathan Seiberg}
\smallskip
\smallskip
\centerline{Department of Physics and Astronomy}
\centerline{Rutgers University }
\centerline{Piscataway, NJ 08855-0849}
\centerline{\tt seiberg@physics.rutgers.edu}
\bigskip
\bigskip

\noindent
We survey the various field theories with 16 real supercharges.  The
most widely known theory in this class is the $N=4$ theory in four
dimensions.  The moduli space of vacua of these theories are
described and the physics at the singularities of the moduli spaces
are studied.

\Date{May 1997}

\newsec{Introduction}

\nref\mo{C. Montonen and D. Olive, \pl {72}{1977}{117}; P. Goddard,
J. Nuyts and D. Olive, \np{125}{1977}{1}.}%
\nref\dualnf{H. Osborn, \pl{83}{1979}{321}; A. Sen, hep-th/9402032,
\pl{329}{1994}{217}; C. Vafa and E. Witten, hep-th/9408074,
\np{432}{1994}{3}.}%

The magic of supersymmetry makes supersymmetric theories amenable to
exact treatment.  With more supercharges, the theory is more constrained
and more observables can be analyzed exactly.  The largest number of
supercharges, which is possible in free field theory, is sixteen.  With
more supercharges the free multiplet includes fields whose spin is
larger than one and no consistent theory (without gravity) exists.
There are three motivations for studying these theories.  First, as the
most supersymmetric theories they are the most constrained theories, and
therefore they exhibit interesting features like exact electric/magnetic
duality in the $N=4$ theory in four dimensions
\refs{\mo, \dualnf}.  Second, these theories appear in string
compactifications as the theory of the collective coordinates of various
branes.  Finally, the $N_c \rightarrow \infty$ limit of these $U(N_c)$
gauge theories have been proposed as exact descriptions of toroidally
compactified M-theory
\ref\bfss{T. Banks, W. Fischler, S.H. Shenker and L. Susskind,
``M Theory as a Matrix Model: A Conjecture,'' hep-th/9610043.}.

In section 2 we survey the various theories in this class and examine
when they have non-trivial infrared dynamics.  In section 3 we review
briefly the $N=4$ theory in four dimensions.  In section 4 we focus on
the $N=8$ theory in three dimensions and study its moduli space of
vacua and its singularities.  Section 5 is devoted to the
compactification of the four dimensional $N=4$ theory on a circle to
three dimensions.  We discuss the image of the famous
electric/magnetic duality of the four dimensional theory in three
dimensions.  In section 6 we make some comments on the $(8,8)$ theory
in two dimensions.  In section 7 we present a few observations on the
special theory with $(0,2)$ supersymmetry in six dimensions.

\newsec{The zoology of theories with 16 supersymmetries}

The most symmetric classical field theory with 16 supersymmetries is
supersymmetric Yang-Mills theory in ten dimensions.  The simplest such
theory is an Abelian gauge theory.  The supermultiplet includes a
massless photon and a massless fermions.  The non-Abelian extension of
this theory exists as a classical field theory but its quantum version
is anomalous and therefore inconsistent. 

The theory in $d$ dimensions, which is obtained by dimensional reduction
of the classical theory in ten dimensions is anomaly free.  Its Lorentz
symmetry is $Spin(d-1,1)$.  The dimensionally reduced theory has an
R-symmetry $Spin(10-d)$, which originates from the ten dimensional
Lorentz
group.  The sixteen generators transform as
\eqn\supertable{\matrix{
d \qquad & \qquad Spin(d-1,1)\times Spin(10-d) \qquad & \qquad {\rm
Automorphism} \supseteq Spin(10-d) \cr
& & \cr
9 \qquad & \qquad {\bf 16}_r \qquad & \qquad \cr
8 \qquad & \qquad ({\bf 8_s,1}) +({\bf \bar 8_s=8_c,-1}) \qquad & \qquad
U(1) =Spin(2) \cr
7 \qquad & \qquad ({\bf 8}_p,{\bf 2}_p) \qquad & \qquad SP(1) = Spin(3)
\cr 
6 \qquad & \qquad ({\bf 4}_p,{\bf 2}_p) +({\bf 4'}_p,{\bf 2'}_p) \qquad
& \qquad SP(1) \times SP(1) = Spin(4) \cr
5 \qquad & \qquad ({\bf 4}_p,{\bf 4}_p) \qquad & \qquad SP(2)=
Spin(5)\cr 
4 \qquad & \qquad ({\bf 2,4}) +({\bf \bar 2, \bar 4}) \qquad & \qquad
U(4)\supset Spin(6)\cr
3 \qquad & \qquad ({\bf 2}_r,{\bf 8}_r)  \qquad & \qquad  Spin(8) \supset
Spin(7) \cr
2 \qquad & \qquad ({\bf 1}_r,{\bf 8_s})+({\bf -1}_r,{\bf 8_c}) \qquad &
\qquad Spin(8) 
\times Spin(8) \supset Spin(8) \cr
1 \qquad & \qquad {\bf 16}_r \qquad & \qquad Spin(16)\supset Spin(9)\cr}}
where the subscripts $p$ and $r$ label pseudoreal and real
representations respectively and we label the three eight dimensional
representations of $spin(8)$ (or its non-compact versions) as ${\bf
8_{s,c,v}}$.  The automorphism group of the supersymmetry 
algebra can be larger than the R-symmetry of the Lagrangian.  It is
determined by the anticommutation relations of the supercharges and
their reality properties ($SP$ for pseudoreal, $Spin$ for real and $U$
for complex).  The automorphism group is also included in the table
along with its $Spin (10-d)$ subgroup.

The only irreducible massless representation, with spin less than one,
of the superalgebra is obtained by dimensional reduction from ten
dimensions.  It includes a photon $A_\mu$, $10-d$ scalars $\phi$, and
some fermions.

One characteristic of all these theories is the existence of a moduli
space of vacua.  The large amount of supersymmetry constrains the moduli
space to be locally flat but there can be singularities.  The moduli
space of a gauge group $G$ of rank $r$ in $d\ge 4 $ dimensions is
\eqn\modulig{\CM= {\BR^{r(10-d)} \over \CW},}
where $\CW$ is the Weyl group of $G$.  For $SU(2)$, $r=1$ and $\CW=\BZ_2$.
The effective Lagrangian along the flat directions is constrained to be
free
\eqn\efflagfl{\CL= {1 \over g^2} \big( F_{\mu\nu}^2 +
(\partial\phi^i)^2 + {\rm fermions} \big)}
where $g$ is the gauge coupling constant.  By supersymmetry it is
independent of $\phi$.

The coupling constant $g$ in \efflagfl\ seems unphysical because the
theory is free and one might attempt to absorb it by rescaling the
dynamical 
variables $\phi$ and $A_\mu$.  However, such a redefinition puts $g$ in
the gauge transformation laws.  Furthermore, when the theory \efflagfl\
is studied on nontrivial manifolds, there exist nontrivial bundles with
magnetic fluxes of the photon.  Their action depends on $g$ and hence
$g$ is physical.

The most interesting aspect of the dynamics of these theories is their
behavior at the singularities of the moduli space.  In a classical
theory based on a non-Abelian gauge theory there are new massless
particles at the singularities.  What happens in the quantum theory?
The standard criterion for nontrivial dynamics in the theory of the
renormalization group is the dimension of the coupling constant.  We
always assign dimension one to gauge fields, $A_\mu$, since their Wilson
line is dimensionless.  By supersymmetry we should also assign dimension
one to the scalars $\phi$.  The gauge coupling $g$ is then of dimension
$4-d\over 2$.  For $d>4$ its dimension is negative, and the
corresponding interaction is irrelevant at long distance.  Therefore, we
do not expect any interesting infrared dynamics in the gauge theory
above four dimensions. 

We can generalize this standard argument and include theories, which
might not come from a Lagrangian but have a moduli space of vacua, where
they are free at long distance.  The long distance theories at the
singularities must be at fixed points of the renormalization group and
therefore they are scale invariant.  These scale invariant theories can
be free (orbifold theories) or interacting.  The effective Lagrangian
along the flat directions is constrained by supersymmetry to have the
form \efflagfl.  Furthermore, since the flat directions emanate from
scale invariant theories this Lagrangian must exhibit {\it spontaneous
breaking of scale invariance}.  However, for $d>4$
\efflagfl\ exhibits explicit breaking of scale invariance.  This follows
from the fact that with $A_\mu$ of dimension one $g$ is dimensionful.
In fact, since the dimension of $g$ is negative, $g$ approaches zero at
long distance.  (Of course, since the theory along the flat directions
is free, we can absorb $g$ in the fields to find a scale invariant
theory.  Such a scaling will not be compatible with a possible
nontrivial theory at the singularities of the moduli space.)  We
conclude that above four dimensions the theories with sixteen
supersymmetries of \supertable, which have a moduli space of vacua
cannot exhibit nontrivial dynamics, and are free at long distance.

The previous argument used only scale invariance and the assumption
about the moduli space of free vacua which emanate from the interacting
point.  Another argument supporting this conclusion is based on the
assumption that all nontrivial fixed points of the renormalization group
with a gap in the spectrum of dimensions of operators are not only scale
invariant but are also conformal invariant\foot{We thank P.  Townsend
and E.  Witten for helpful discussions on this point.}.  In a
supersymmetric theory they should also be superconformal invariant.  The
possible superconformal algebras were analyzed in
\ref\nahm{W.Nahm, ``Supersymmetries and Their Representations,'' \np
{135}{1978}{149}.} 
with the conclusion that the supersymmetry algebras in \supertable\ do
not admit an extension to a superconformal algebra above four
dimensions.  We should note, though, that the free theory is scale
invariant but not conformal invariant.  This comes about by
absorbing the coupling constant $g$ in the photon.  Then the photon
field $A_\mu$ does not have dimension one and the gauge invariant field
strength $F_{\mu\nu}$ does not have dimension two.  It is expected that
such scale invariance without conformal invariance does not extend to
interacting theories.

The most widely known theory in this class is the $N=4$ theory in four
dimensions.  The scaling argument shows that it can be scale
invariant -- in this case the Lagrangian \efflagfl\ along the flat
directions is scale invariant.  Even if higher derivative terms are
included 
\ref\dinesei{M. Dine and N. Seiberg, ``Comments on Higher Derivative
Operators in Some SUSY Field Theories,'' hep-th/9705057, SCIPP 97/12,
RU-97-31 }, 
the effective Lagrangian exhibit spontaneous breaking of scale
invariance.  Indeed, it is known that the Yang-Mills theory is a finite,
superconformal field theory.  We will analyze some of its properties in
section 3.

Below four dimensions $g$ has positive dimension.  Therefore, it
corresponds to a relevant operator in the Lagrangian, and the
long distance behavior can be different from the short distance
description.  In sections 4, 5 and 6 we will analyze the three
dimensional and the two dimensional theories.

There is one more supersymmetry algebra with sixteen supercharges, which
is not included in \supertable.  It is in six dimensions and includes
four spinors of the same chirality.  It is usually called the $(0,2)$
algebra, while the one in \supertable\ in six dimensions is the $(1,1)$
algebra.  Its automorphism group is $SP(2)$ and the supercharges are in
$({\bf 4}_p, {\bf 4}_p)$ of the $Spin(5,1) \times SP(2)$.  Its
irreducible massless representation consists of a two-form $B_{\mu\nu}$,
whose field strength $H=dB$ is selfdual, five real scalars $\Phi$ and
fermions.  Upon compactification to lower dimensions this theory becomes
one of the theories in \supertable.

Despite recent successes
\ref\recent{P. Pasti, D. Sorokin and M. Tonin, ``Covariant Action for a
D=11 Five-Brane with the Chiral Field,'' hep-th/9701037;  M. Aganagic,
J. Park, C. Popescu and J.H. Schwarz, ``World-Volume Action of the M
Theory Five-Brane,'' hep-th/9701166.}, 
no fully satisfactory Lorentz invariant Lagrangian for a two form with
selfdual field strength is known.  Ignoring the self-duality constraint
the free Lagrangian is
\eqn\efflagfla{H_{\mu\nu\rho}^2 +  (\partial\Phi^i)^2 + {\rm fermions};}
i.e.\ the metric on the moduli space is locally flat.  Note that for $H$
to be selfdual, there cannot be an arbitrary coupling constant $g$ in
front of this Lagrangian.

Let us repeat our analysis of scale invariance.  We should assign
dimension two to the two form $B_{\mu\nu}$ and hence, by supersymmetry
we should assign dimension two to $\Phi$.  Hence, \efflagfla\ is scale
invariant.  Therefore, if there are singularities in the moduli space,
the theory there can be a nontrivial field theory.  Indeed, the analysis
of \nahm\ shows that the $(0,2)$ supersymmetry algebra admits an
extension to the superconformal algebra.  The $SP(2)$ automorphism group
of the supersymmetry algebra is included in the superconformal algebra.
In section 7 we will make some comments on this theory.

\nref\seibergdfive{N. Seiberg, ``Five Dimensional SUSY Field Theories,
Non-trivial Fixed Points and String Dynamics,'' \pl {388} {1996}
{753}, hep-th/9608111.}%
\nref\IMS{K. Intriligator, D. Morrison and N. Seiberg,
``Five-Dimensional Supersymmetric Gauge Theories and Degenerations of 
Calabi-Yau Spaces,'' hep-th/9702198,  RU-96-99, IASSNS-HEP-96/112.}%
\nref\ganoha{O. Ganor and A. Hanany, ``Small E(8) Instantons and
Tensionless Noncritical Strings,'' hep-th/9602120.}%
\nref\sws{N. Seiberg and E. Witten, ``Comments on String Dynamics in
Six Dimensions,'' hep-th/9603003, \np{471}{1996}{121}.}%
\nref\dlp{M. Duff, H. Lu and C.N. Pope, ``Heterotic Phase
Transitions and Singularities of The Gauge Dyonic String,''
CTP-TAMU-9-96, hep-th/9603037, \pl{378}{1996}{101}.}%
\nref\seibergds{N. Seiberg, ``Non-trivial Fixed Points of The
Renormalization Group in Six Dimensions,'' \pl {390}{1997}{169},
hep-th/9609161,}%
\nref\bv{M. Bershadsky and C. Vafa, ``Global Anomalies and Geometric
Engineering of Critical Theories in Six Dimensions,'' hep-th/9703167.}%

Although in these notes we focus on theories with 16 supercharges, we
would like to mention a similar analysis of theories with 8
supercharges above 4 dimensions.  In certain examples in 5 dimensions 
\refs{\seibergdfive, \IMS} the effective Lagrangian along the flat
directions 
\eqn\effdfive{\phi F_{\mu\nu}^2 + \phi (\partial \phi)^2 + ...}
exhibits spontaneous breaking of scale invariance (as before, the
dimension of $A_\mu$ and of $\phi$ is one) and the strongly coupled
theory at the singularity is scale invariant.  Similarly, in six
dimensions, the tensor multiplet includes a scalar $\Phi$ and a two
form $B_{\mu\nu}$ both of dimension 2 and hence the Lagrangian
\eqn\effdsixe{\Phi F_{\mu\nu}^2 + B \wedge F \wedge F + (\partial
\Phi)^2 + (dB)^2 ...}
is scale invariant and the singularities in the moduli space can be
strongly coupled scale invariant theories
\refs{\ganoha - \bv}
Again, this is consistent with the existence of a superconformal
extension of $N=1$ supersymmetry in five and six dimensions \nahm.

\newsec{$N=4$ supersymmetry in $d=4$}

The superconformal algebra includes an $SU(4)=Spin(6)$ symmetry \nahm.
The extra $U(1)$ factor in the automorphism group of the supersymmetry
algebra (see \supertable) is not a symmetry.  This theory is labeled
by a dimensionless coupling constant $g$.  One can also add the theta
angle to make it complex
\eqn\deftau{\tau= {\theta \over 2\pi} + {4 \pi \over g^2} i.}

The moduli space of vacua is, as in \modulig, $\BR^{6r} /\CW$.  Since the
theory is scale invariant, the expectation values of the scalars along
the flat directions lead to spontaneous scale symmetry breaking.  One of
the massless scalars is a dilaton -- the Goldstone boson of broken scale
invariance.  At the generic point in the moduli space the low energy
theory is $U(1)^r$ and the global $Spin(6)$ symmetry is spontaneously
broken.  However, at long distances this theory is free and
a new $Spin(6)$ symmetry appears.  This is consistent with the fact that
at long distance we find a conformal field theory and the $Spin(6)$
R-symmetry is included in the conformal algebra.

At the singularities some of the gauge symmetry is restored.  More
precisely, the theory at the singularities is an interacting conformal
field theory.  In such a theory the notion of particles is ill defined,
and in particular, we cannot say that the gauge symmetry is restored
because the gauge bosons are meaningless.  It is standard to use the
superconformal symmetry to analyze the theory there.  Primary
operators are defined to be the operators which are annihilated by all
the superconformal generators.  The (anti)commutation relations lead
to a bound on their dimensions  $D(\CO)$ in terms of their $Spin(6)$
representations.  Polynomials in the microscopic fields, which are
scalars of the Lorentz group, must be in representations of $SO(6)$.
Some examples of the inequality for their dimensions are
\eqn\nfourbo{\eqalign{
&D ( {\bf 6}) \ge 1 \cr
&D ( {\bf 10}) \ge 3 \cr
&D ( {\bf 15}) \ge 2 \cr
&D ( {\bf 20'}) \ge 2 \cr
}}
The inequality \nfourbo\ is saturated for
chiral fields.  Indeed, along the flat directions, where we find a free
field representation of the algebra the dimension of the free scalar
field in the $\bf 6$ of $Spin(6)$ is one.  Two bosons are in ${\bf 6}
\times {\bf 6} = {\bf 1}_s + {\bf 15}_a + {\bf 20'}_s$.  The operator
in the singlet can mix with the identity operator and is not chiral.
The ${\bf 15}$ occurs only when we have more than one scalar.  Then
there is no short distance singularity in forming the composite field in
${\bf 15}$, and its dimension is clearly 2, which is consistent with
\nfourbo.  Finally, the 
fact that the dimension of the composite field ${\bf 20'}_s$ is 2
follows from examining an $N=1$ superconformal subalgebra and noticing
that it includes a field with $R={4 \over 3} $.

At the origin the theory is interacting but we can still use \nfourbo\
to determine the dimensions of gauge invariant chiral operators.  For
example, the dimension of the scalar bilinear $\Tr \phi^i \phi^j$ in
${\bf 20'}_s$ of $Spin(6)$ is 2.  Note that the dimensions of chiral
operators are independent of $\tau$ -- they are given by their value
in free field theory.  This is not the case for more general
operators.  The theory at the origin must have a truly marginal operator
corresponding to changing the value of $\tau$.  It seems to be the
$Spin (6)$ invariant
$Q^2 \Tr (\phi^i \phi^j \phi^k)_{\bf 10} \sim Q^4 \Tr (\phi^i
\phi^j)_{\bf 20'} $
(the equality of these expressions seems to follow from the equation of
motion). 

This theory is expected to exhibit electric/magnetic duality \refs{\mo,
\dualnf} -- the theory characterized by the gauge group $G$, whose weight
lattice is $\Gamma_w(G)$ and coupling constant $\tau$ is the same as the
theory based on the dual gauge group ${}^*G$, whose weight lattice is
\eqn\dualgro{\Gamma_w({}^*G)={}^*\Gamma_w(G)}
and coupling $-1/\tau$.  The spectrum includes BPS particles with
electric and magnetic charges.  Since this duality and its action on the
spectrum are well known, we will not review it here. 

Some of the higher dimension operators along the flat directions, which
correct the leading order terms \efflagfl, were analyzed in \dinesei.
The leading irrelevant operator is of the form
\eqn\leadirr{ {1 \over \phi^4} F^4 + {1 \over \phi^4} (\partial \phi)^4
+ {\rm eight ~ fermion ~ terms}.} 
Supersymmetry leads to a non-renormalization theorem guaranteeing that
these terms are generated only at one loop and are not corrected by
higher order perturbative or nonperturbative effects.  We will not
repeat the argument here and refer the reader to \dinesei.

\newsec{$N=8$ supersymmetry in $d=3$}

Here we study field theories with $N=8$ supersymmetry in $d=3$.  The
super generators are in the real two dimensional representation of the
Lorentz group.  The automorphism group of the algebra (R-symmetry) is
$Spin(8)$ and the supergenerators transform as an eight dimensional
representation, which we take to be the spinor $\bf 8_s$.

Since for massless particles the little group is trivial, there is only
one massless representation of the superalgebra.  It consists of 8
bosons in $\bf 8_v$ and 8 fermions in $\bf 8_c$ of $Spin(8)$.  Starting
with a higher dimensional field theory with the same number of
supersymmetries (e.g.\ $N=4$ in $d=4$) we find a vector field, 7 scalars
and 8 fermions.  The R-symmetry, which is manifest in this description,
is $Spin(7) \subset Spin(8)$.  The vector is a singlet of $Spin(7)$,
the scalars are in $\bf 7$ and the fermions in $\bf 8$.  After
performing a duality transformation on the vector it becomes a scalar
and the $Spin(8)$ symmetry becomes manifest.

Interacting Lagrangians with $N=8$ supersymmetry do not necessarily
exhibit the maximal possible R-symmetry.  In particular, the Yang-Mills
Lagrangian is invariant only under the $Spin(7)$ subgroup.  At long
distance, the theory must flow to a scale invariant theory, which we
assume to be also superconformal invariant.  The conformal algebra in 3
dimensions is $Spin(3,2)$.  The sixteen supersymmetry generators combine
with sixteen superconformal generators to eight spinors of $Spin(3,2)$.
For the closure of the algebra we must include the the $Spin(8)$
symmetry \nahm.  Hence, the long distance theory is invariant under the
full R-symmetry $Spin(8)$.  More generally, with $N$ supercharges the
superconformal algebra includes a $Spin(N)$ R-symmetry under which the
supercharges transform as a vector.  For $N=2$ supersymmetry the
R-symmetry is $U(1)$, and normalizing the charge of the supercharge to
be one, we have for scalar operators $D \ge R$.  For $N=4$ supersymmetry
the R-symmetry is $SU(2)_1 \times SU(2)_2$ with the supercharges in the
vector $(I_1= \half, I_2=\half)$ and for scalar fields $D \ge I_1 +
I_2$.  For $N=8$ we use the triality of $Spin(8)$ to put the
supercharges in $\bf 8_s$ rather than in a vector. This leads to the
bounds on the dimensions
\eqn\dimenei{\eqalign{
&D({\bf 8_v}) \ge {1 \over 2} \cr
&D({\bf 8_s}) \ge 1 \cr
&D({\bf 28}) \ge 1 \cr
&D({\bf 56_v}) \ge {3 \over 2} \cr
&D({\bf 35_s}) \ge 2 \cr
&D({\bf 35_v}) \ge 1 \cr
}}
and the bound on $D({\bf r_c})$ the same as for $D({\bf r_v})$.

As an example, consider the $N=8$ supersymmetric $SU(2)$ gauge theory.
The gauge coupling $g$ has dimension $\half$, and therefore the theory
is superrenormalizable.  To analyze its long distance behavior we start
by considering the moduli space of vacua.  Along the flat directions the
$SU(2)$ gauge symmetry is broken to $U(1)$.  The low energy degrees of
freedom are in a single $N=8$ multiplet, which includes seven scalars
$\phi^i$ ($i=1,...,7$) and a photon.  Their Lagrangian is as in
\efflagfl 
\eqn\efflag{ { 1\over g^2} (F_{\mu\nu}^2 + (\partial \phi^i)^2).}
The dual of the photon is a compact scalar $\phi^0$ of radius one with
the Lagrangian
\eqn\duallag{{ 1\over g^2} (\partial \phi^i)^2 + g^2 (\partial
\phi^0)^2.} 
Because of $N=8$ supersymmetry, the leading terms in the Lagrangian
\duallag\ are not corrected 
in the quantum theory.  Therefore, the moduli space of vacua $\CM$ is
eight real dimensional.  The $\phi^i$ label $\BR^7$ and $\phi^0$ labels
$\bS^1$.  The Weyl group of $SU(2)$ changes the sign of $(\phi^i,
\phi^0)$ and therefore 
\eqn\modsp{\CM={\BR^7\times \bS^1 \over\BZ_2} .}
It has two singularities at $\phi^i=\phi^0=0$ and at $\phi^i=0$,
$\phi^0=\pi$.  The metric around them is an orbifold metric.

At long distance the gauge coupling $g$ goes to infinity.  The radius of
the circle in \modsp\ goes to infinity and we can focus on a
neighborhood in the moduli space.  At the generic point we find a free
field theory.  The theory at the two orbifold singularities is more
interesting.  The moduli space around each of them looks like
$\BR^8/\BZ_2$.  We will soon argue that the singularity at $\phi^i=0$,
$\phi^0=\pi$ is simply an orbifold singularity -- the theory at this
point is a free field theory with a gauged $\BZ_2$ symmetry.  The other
singularity, at $\phi^i=0$, $\phi^0=0$ is likely to be an interacting
superconformal field theory.

\nref\sethisu{S. Sethi, L.Susskind, ``Rotational Invariance in the
M(atrix) Formulation of Type IIB Theory,'' hep-th/9702101.}%
\nref\bansei{T. Banks and N. Seiberg, ``Strings from Matrices,''
hep-th/9702187, RU-97-6.}%

Along the flat directions we get from the eight bosons, which are the
fluctuations around the expectation values of $\phi^i$ and $\phi^0$, an
$\bf 8_v$ of the $Spin(8)$ R-symmetry.  Equation \dimenei\ shows that
the dimension of the bosons is $\half$.  This is exactly the result in a
free field theory.  We will show that at the singularity at $\phi^i=0$,
$\phi^0=\pi$ the theory is free.  Since it is an orbifold theory, the
fluctuations $\phi^i$ and $\phi^0$ are not gauge invariant.  Only
bilinears in them are gauge invariant operators.  Their dimensions are
determined easily using \dimenei.  The interacting theory at
$\phi^i=\phi^0=0$ is more interesting.   It will be interesting to find
the leading irrelevant operator there (it seems to be $Q^4 \CO_{\bf
35_s}$).  The $Spin(8)$ invariance of this theory was crucial 
in a recent discussion of the Matrix model applications of this theory
\refs{\sethisu, \bansei}.

More generally consider a gauge group $G$ with rank $r$.  Along the flat
directions $G$ is broken to its Cartan torus $\bT(G) =
\BR^r/{}^*\Gamma_w(G)$, where $\Gamma_w(G)$ is the weight lattice of the
group $G$ and ${}^*\Gamma_w(G)$ is its dual.  The $r$ photons can be
dualized to $r$ scalars taking values in
$\BR^r/\Gamma_w(G)=\BR^r/{}^*\Gamma_w({}^*G)=\bT({}^*G) $, where ${}^*G$
is the dual group, whose weight lattice is dual to $\Gamma_w(G)$.
Therefore, the moduli space is
\eqn\generalmod{\CM(G) = {\BR^{7r} \times \bT({}^*G) \over \CW},}
where $\CW$ is the Weyl group of $G$.

For example, let us compare the $SU(2)$ gauge theory with its dual group
$SO(3)=SU(2)/\BZ_2$.  Since $\bT(SU(2)) /\BZ_2 =\bT(SO(3))$,
\eqn\sutsot{\CM(SU(2)) = {\CM(SO(3)) \over \BZ_2}.}
The $SO(3)$ gauge theory has a global $\BZ_2$ symmetry, which shifts
$\phi^0$ by half its periodicity.  In the $SU(2)$ theory this $\BZ_2$
symmetry becomes a gauge symmetry and the moduli space is modded out by
it\foot{There is a similar symmetry in the analogous $N=4$ theory in
three dimensions.  The moduli space of the $SU(2)$ gauge theory
was determined in
\ref\compacti{N. Seiberg and E. Witten,``Gauge Dynamics
and Compactification to Three Dimensions,'' IASSNS-HEP-96-78,
hep-th/9607163.} 
to be the Atiyah-Hitchin space.  Its fundamental group is $\BZ_2$.  If
we instead consider the $SO(3)$ theory, the moduli space becomes the
double cover of the Atiyah-Hitchin space, and the $\BZ_2$ is a global
symmetry.  This fact is in accord with the discussion of confinement in
\compacti.  In the $SU(2)$ theory there is (with a suitable
perturbation) confinement of electric charge modulo 2 -- the massive W
bosons can screen external sources.  This is reflected in the
fundamental group of the moduli space being $\BZ_2$.  In the $SO(3)$
theory there are no integer external sources and therefore there is no
confinement.  Correspondingly, the fundamental group of the moduli space
is trivial.}.

Such three dimensional gauge theories are realized in the study of
D2-branes
\ref\joe{For reviews, see, J. Polchinski, S. Chaudhuri, and C. Johnson, 
``Notes on D-branes,'' hep-th/9602052; J. Polchinski, ``TASI Lectures on
D-branes,'' hep-th/9611050.}
in the IIA theory in ten dimensions.  The collective coordinates of
every D2-brane form a single vector multiplet
\ref\collec{E. Witten, ``Bound States of Strings and p-Branes,''
\np{460}{1996}{335}, hep-th/9510135.}.
The 7 scalars correspond to the 7 transverse directions of the brane.
The dual of the vector multiplet corresponds to the position of the
brane in the eleventh compact dimension
\ref\townsend{P. Townsend, ``D-Branes From M-Branes,''
\pl{373}{1996}{68}, hep-th/9512062.}.
Hence the moduli space of vacua of the D2-brane is $\BR^7 \times \bS^1$.
The coupling constant of this three dimensional field theory determines
the circumference of the $\bS^1$ factor, such that in the strong
coupling limit the radius goes to infinity and the membrane propagates
in flat eleven dimensional space.

A configuration of two D2-branes in the IIA theory is described by a
$U(2)$ gauge theory \collec.  At the generic point in the moduli space
of vacua the $U(2)$ gauge symmetry is broken to $U(1)^2$ and the light
fields are two vector multiplets of $N=8$.  After dualizing the photons
and modding out the the Weyl group (which interchanges the two $U(1)$
factors) we find the moduli space
\eqn\modspacetwo{\CM_2 = {(\BR^7 \times \bS^1) \times (\BR^7 \times
\bS^1) \over \BZ_2},}
which is labeled in an obvious way by $\phi^i_I$ and $\phi^0_I$
($I=1,2$) and the $\BZ_2$ interchanges $I=1$ with $I=2$.  The
singularities in $\CM_2$ are at $\phi^i_1= \phi^i_2$ and $\phi^0_1=
\phi^0_2$.  Physically, they occur when the two membranes are on top of
each other.

What is the physics at this singularity?  At short distance all the
degrees of freedom of the $U(2)$ gauge theory are physical.  The
interactions between them become strong as we approach the infrared.
At long distance the theory flows to a superconformal field theory.
This theory may be an interacting or a free field theory.  Either way
the degrees of freedom at long distance differ from the degrees of
freedom at short distance.  If the theory is interacting, the notion
of the particles at long distance is ill defined.  If, however, the
theory there is free, it includes only {\it two} supermultiplets
(rather than the four supermultiplets of the UV $U(2)$ gauge theory).
Although we cannot prove it, we find it more likely that the theory
there is actually interacting\foot{The Matrix model description of IIB
strings is based on this 2+1 dimensional fixed point field theory
\refs{\sethisu, \bansei}.  Nontrivial string interactions in this
framework arise only if this fixed point is interacting.}.  As we
argued above, this interacting theory has $Spin(8)$ enhanced symmetry.

Consider now modding out this theory by the common ``center of mass
motion'' to derive an $SU(2)$ gauge theory.  The moduli space of vacua
is $\CM= (\BR^7\times \bS^1)/ \BZ_2$, where the $\BR^7$ is parametrized
by $\phi^i=\phi^i_1-\phi^i_2$ and the $\bS^1$ is parametrized by
$\phi^0=\phi^0_1-\phi^0_2$. The singularity at $\phi^i=\phi^0=0$ is the
same as in the $U(2)$ problem and is likely to be interacting.  What
about the other singularity at $\phi^i=0$, $\phi^0=\pi$?  It arises
because the notion of center of mass in the $\bS^1$ direction is ill
defined.  The moduli space of vacua has a new singularity when the two
membranes are at $\phi^i_1=\phi^i_2$ and $\phi^0_1=\phi^0_2+\pi$; i.e.\
when the two membranes are at the same point in $\BR^7$ but at antipodal
points in the $\bS^1$.  Clearly, the dynamics at this point is trivial
and the singularity is a consequence of the fact that we change $U(2)$
to $SU(2)$ -- this is merely an orbifold singularity.  Therefore, the
singularity at $\phi^i=0$, $ \phi^0=\pi$ in the $SU(2)$ theory is an
orbifold singularity and the theory there is free.

We remarked above that the leading irrelevant operators along the flat
directions of the corresponding four dimensional theory are subject to a
non-renormalization theorem and are generated only at one loop.  The
analogous theorem does not hold in three dimensions \dinesei.  The
classical static configuration of the 'tHooft-Polyakov monopole appears
as an instanton in the three dimensional theory.  One of the effects of
these instantons is to explicitly break the shift symmetry of the
various magnetic photons $\phi^0$.  In fact, we have already mentioned
that although the metric on the moduli space \generalmod\ is flat and
hence invariant under the shift, the singularities are not invariant.
These instantons were first studied in the theory without supersymmetry
by Polyakov
\ref\polyakov{A.M. Polyakov,``Quark Confinement And The Topology Of Gauge 
Groups,'' \np{120}{1977}{429}.}.
In theories with $N=2$ supersymmetry they were discussed by Affleck,
Harvey and Witten
\ref\ahw{I. Affleck, J. Harvey and E. Witten, ``Instantons And
(Super)Symmetry Breaking In $2+1$ Dimensions,''\np {206}{1982}{413}.}
and in $N=4$ in
\nref\swthree{N. Seiberg and E. Witten,``Gauge Dynamics
and Compactification to Three Dimensions,
`` IASSNS-HEP-96-78, hep-th/9607163.}%
\nref\losalamosf{N. Dorey, V.V. Khoze, M.P. Mattis, D. Tong, and S.
Vandoren, ``Instantons,
Three-Dimensional Gauge Theory, and the Atiyah-Hitchin Manifold,''
hep-th/9703228.}%
\nref\polpou{J. Polchinski and P. Pouliot,  ``Membrane Scattering with M
Momentum Transfer,'' NSF-ITP-97-27,hep-th/9704029.}%
\nref\dkm{N. Dorey, V.V. Khoze and M.P. Mattis, ``Multi-Instantons,
Three-Dimensional Gauge Theory, and the Gauss-Bonnet-Chern Theorem,''
hep-th/9704197.}%
\refs{\swthree,\losalamosf}.  Their effects in the $N=8$ theory were
studied in \refs{\polpou,\dkm,\dinesei}. In particular, they were shown
to contribute to terms with four derivatives and to terms with eight
fermions (as well as to other terms).   More effects of these terms will
be discussed in
\ref\bfssn{T. Banks, W. Fischler, N. Seiberg and L. Susskind, to
appear.}. 

\newsec{The $N=4$, $d=4$ theory on $\BR^3 \times \bS^1$}

Consider now starting with a higher dimensional theory with 16
supercharges and compactifying on a torus to three dimensions.  Some
of the scalars in the three dimensional Lagrangian originate from
components of gauge fields in the higher dimensional theory.
Therefore, the corresponding directions in the moduli space of the
three dimensional theory must be compact.  Let us start by considering
the free $U(1)$ $N=4$ theory in $d=4$ with gauge coupling $g_4$ and
compactify it on a circle of radius $R$ to three dimensions.  The
three dimensional gauge coupling $g_3$ satisfies
\eqn\threedgac{ {1 \over g_3^2} = {R \over g_4^2}.}
The six scalars in the vector multiplet in four dimensions become
$\phi^i$ with $i=1,...,6$.  $\phi^7 $ arises from a component of the
four dimensional gauge field $\phi^7=A_4$.  It corresponds to a $U(1)$
Wilson line around the circle.  A gauge transformation, which winds
around this circle, identifies $\phi^7 $ with $\phi^7 + {1 \over R}$.
Therefore, we define the dimensionless field $\phi_e= RA_4$, whose
circumference is one.  When we dualize the three dimensional photon to
a scalar $\phi_m$, we find the Lagrangian \compacti
\eqn\fourthreelag{{R \over g_4^2}(\partial \phi^i)^2 +{1 \over
Rg_4^2}(\partial \phi_e)^2 + {g_4^2 \over R} (\partial \phi_m)^2.}
The moduli space of vacua is 
\eqn\fourtothree{\BR^6 \times \bT^2 }
where the two circles in $\bT^2$ correspond to the two compact bosons
$\phi_e$ and $\phi_m$.  They represent a $U(1)$ Wilson line and a $U(1)$
'tHooft line around the circle we compactified on. In other words, these
two scalars are the fourth component of the $d=4$ photon $A_4$ and the
fourth component of the magnetic photon $\tilde A_4$.  The non-trivial
duality transformation in $d=4$ is translated to
\eqn\fourddul{\eqalign{
\phi_e \rightarrow \phi_m \cr
\phi_m \rightarrow - \phi_e \cr
g_4 \rightarrow {1 \over g_4} .\cr}}
It is easy to add the $\theta$ angle in
four dimensions and recover the $SL(2,\BZ)$ action in four dimension as
an action on the $\bT^2$ in the moduli space \fourtothree.

As we said above, at long distance in the three dimensional theory
only the local structure of the moduli space \fourtothree\ matters.
It is $\BR^8$.  The eight scalars transform as a vector under the
enhanced $Spin(8)_R$ symmetry.  The duality transformation \fourddul\
becomes part of the $Spin(8)_R$ symmetry.

We can easily extend this discussion to compactified interacting
theories.  For example, consider the $SU(2)$ $N=4$ theory in $d=4$.
Repeating the analysis of the $U(1)$ theory and modding out by the Weyl
group, we find the moduli space of vacua
\eqn\fourtothreet{{\BR^6 \times \bT^2 \over \BZ_2}.}
The moduli space has four orbifold singularities.  As before, the
theory at three of them are orbifold theories (the metric at all of
them is an orbifold metric) and the fourth is likely to be an
interacting superconformal field theory.

The full theory is invariant only under the $Spin(6)$ symmetry of the
four dimensional theory.  The $SL(2,\BZ)$ duality is {\it not} a
symmetry of the theory.  It relates theories with different values of
the coupling constant.  After the compactification this $SL(2,\BZ)$
acts on the $\bT^2$ factor.  Again, it is not a symmetry.  However, at
long distance its $\BZ_2$ subgroup \fourddul\ becomes a symmetry.
Therefore, the symmetry at long distance includes $Spin(6)\times
\BZ_2$.  The three dimensional Lagrangian is obtained by shrinking the
compactification radius $R$ with $g_3$ fixed.  Then, the $Spin(6)$
R-symmetry of the four dimensional theory is enhanced to $Spin(7)$, 
which is manifest in the three dimensional Lagrangian.  Since in this
limit $g_4 \rightarrow 0$, the $\BZ_2$ subgroup of $SL(2,\BZ)$ is not
visible. In the long distance limit we should find a symmetry, which
includes both this $Spin(7)$ R-symmetry and $Spin(6)\times \BZ_2$.
This must be $Spin(8)$.  This leads to an independent derivation of
the $Spin(8)$ symmetry of the long distance theory (the other
derivation was based on its superconformal invariance).  This derivation
was also given in \sethisu.

We conclude that the electric-magnetic duality of the four dimensional
theory becomes a symmetry of the three dimensional theory.  It is
included in its $Spin(8)_R$ R-symmetry.

We now generalize to compactification of a $d=4$, $N=4$
gauge theory with gauge group $G$ of rank $r$ on a circle of radius
$R$ to three dimensions.  Along the flat directions $G$ is broken to
its Cartan torus $\bT(G) =\BR^r/{}^* \Gamma_w(G)$.  The Wilson lines
around the circle lead to $r$ scalars in $\bT(G)$, whose scale is $1
\over \sqrt{R} g_4$.  The 'tHooft loops around the circle (or
equivalently the dual of the three dimensional photons) lead to $r$
scalars on $\bT({}^*G)$, whose scale is $g_4 \over \sqrt{R}$.  The
total moduli space is therefore
\eqn\totalmodft{\BR^{6r} \times \bT(G) \times \bT({}^*G) \over \CW}
where again $\CW$ is the Weyl group.  In this form it is clear that
electric-magnetic duality exchanges $g_4$ with its inverse and $G$ with
${}^*G$.

\newsec{$(8,8)$ supersymmetry in $d=2$}

The $(8,8)$ supersymmetry algebra in two dimensions has a simple
massless free field representation consisting of 8 bosons $\phi^i$, 8
left moving fermions $S_-$ and 8 right moving fermions $S_+$.  It is
interesting to consider the possible action of the $Spin(8) \times
Spin(8)$ automorphism group of the algebra \supertable.  Without loss of
generality, let the 8 right moving supercharges, $Q_+^{\dot \alpha}$, be
in $\bf 8_s$ and the 8 right moving fermions, $S_+^\alpha$, in $\bf 8_c$
of one of the $Spin(8)$ factors.  This implies that the 8 bosons are in
$\bf 8_v$.  Since the bosons are rotated by this symmetry, the
corresponding conserved currents $j_\mu^{[ij]}= \phi^{[i} \partial_\mu
\phi^{j]}$ are not conformal fields -- $\phi^i$ (and not only their
derivatives) appear in the current.  Therefore, we do not have separate
left moving and right moving currents.  This means that the same
$Spin(8)$ symmetry must also act on the left moving supercharges $Q_-$
and fermions $S_-$.  Here we have two options: the left moving
supercharge, $Q_-^{\alpha}$, can be in $\bf 8_c$ and the left moving
fermion,
$S_-^{\dot \alpha}$, in $\bf 8_s$, or the left moving supercharge,
$Q_-^{\dot \alpha}$, can be in $\bf 8_s$ and the left moving fermion,
$S_-^\alpha$, in $\bf 8_c$.  These two assignments appear in the 
Green-Schwarz light cone formalism for IIA and IIB superstrings
respectively.

Interacting theories can be constructed by starting with super
Yang-Mills theory in higher dimensions.  These theories have a global
$Spin(8)$ symmetry.  The left moving and right moving supercharges have
opposite $Spin(8)$ chirality and hence this corresponds to the first
option above (as in IIA strings).  Classically these theories have a
moduli space of vacua \modulig\ $\BR^{8r} /\CW$ ($r$ is the rank of the
gauge group and $\CW$ is its Weyl group).  Along the flat directions the
massless spectrum consists of $r$ copies of the free representation
discussed above with the global $Spin(8)$ symmetry acting as in the
first assignment. However, in two dimensional field theory the notion of
moduli space of vacua is ill defined because we should integrate over
it.  To define it we can integrate out the high energy modes and
construct an effective action for the light modes.  At short distance
the theory is the non-Abelian gauge theory. 
The long distance theory is a scale invariant theory.  Its target space
is the orbifold
\ref\hms{J. Harvey, G. Moore and A. Strominger, ``Reducing S Duality to
T Duality,'' \physrev{52}{1995}{7161}, hep-th/9501022;
M. Bershadsky, A. Johansen, V. Sadov and C. Vafa, ``Topological
Reduction of 4-D SYM to 2-D Sigma Models,'' \np{448}{1995}{166},
hep-th/9501096.} 
\eqn\orbi{{\BR^{8r} \over \CW }.}
The main question is how to treat the theory at the singularities.

In order to answer this question we could attempt to extend the scale
invariance of the theory to conformal invariance and construct a
superconformal field theory with $(8,8)$ supersymmetry.  However,
there is no superconformal extension of this $(8,8)$ supersymmetry
algebra\foot{We thank N. Berkovits for a useful discussion on this
point.} \nahm.  One way to see that is to recall the fact,
demonstrated above in the free representation, that the $Spin(8)
\times Spin(8)$ automorphism group cannot be a symmetry.  Furthermore,
even the currents of the diagonal $Spin(8)$, which can be conserved, are
not conformal fields.  This suggests that perhaps the only scale
invariant theories with $(8,8)$ supersymmetry are free.  In this case,
the long distance theory is simply the orbifold theory based on the
orbifold \orbi. 

In a beautiful paper Dijkgraaf, Verlinde and Verlinde
\ref\dvv{R. Dijkgraaf, E. Verlinde and H. Verlinde, ``Matrix String
Theory,'' hep-th/970330.} analyzed this orbifold conformal field theory
and determined the leading irrelevant operator in the long distance
orbifold theory.  For the simple case of $SU(2)$ it is constructed as
follows.  The target space is $\BR^{8} /\BZ_2$ where the $\BZ_2$ acts by
changing the sign of the 8 bosons $\phi^i$, the 8 right moving fermions
$S_+^\alpha$ and the 8 right moving fermions $S_-^{\dot \alpha}$.  The
right moving bosonic twist operator $\sigma$ and the right moving
fermionic twist fields $\Sigma^i$ and $\Sigma^{\dot \alpha}$ satisfy
\eqn\bostw{\eqalign{&\partial_+  \phi^i(z) \sigma(0) \sim {1 \over
\sqrt {z}} \tau^i(0) \cr
&S_+^\alpha(z) \Sigma^i(0) \sim {1 \over \sqrt {z}} \gamma^i_{\alpha
\dot \alpha} \Sigma^{\dot \alpha} (0) \cr 
&S_+^\alpha(z) \Sigma^{\dot \alpha} (0) \sim {1 \over \sqrt {z}}
\gamma^i_{\alpha \dot \alpha} \Sigma^i (0) .\cr}}
The dimensions of $\sigma$, $\Sigma^i$ and $\Sigma^{\dot \alpha}$ are
$\half$ and the dimension of $\tau^i$ is $1$.  Using these building
blocks Dijkgraaf, Verlinde and Verlinde construct the primary field
$\CO^{\dot \alpha} = \sigma \Sigma^{\dot \alpha}$.  Its ``descendants''
$Q_+^{\dot \alpha}\CO^{\dot \beta}$ can be in ${\bf 1} \oplus {\bf 28}
\oplus {\bf 35_s}$.  An explicit computation shows that the field in
${\bf 35_s}$ is null 
\eqn\nulla{Q_+^{\dot \alpha}\CO^{\dot \beta} + Q_+^{\dot
\beta}\CO^{\dot \alpha} =0.} 
The field in ${\bf 1} $
\eqn\irrop{\CO_+ = \tau^i \Sigma^i}
satisfies, by using \nulla\ and the anticommutation relations
\eqn\susyinv{Q_+^{\dot \alpha}\CO_+ = \partial_+(\sigma \Sigma^{\dot
\alpha}).}
Note that in establishing \susyinv\ we use only the supersymmetry
algebra and not the non-existing superconformal algebra.  We can repeat
this analysis for the left movers and construct the operator $\CO_-$.
Then, because of \susyinv\ the operator
\eqn\finirre{V=\int \CO_+\CO_-}
is supersymmetric.  Its dimension is $({3 \over 2}, {3 \over 2})$ and
it is the leading supersymmetric irrelevant operator. 

As in the previous section we can consider the compactification of the
four dimensional theory on $\bT^2$ to two dimensions and study the
theory as a function of the coupling constant of the four dimensional
theory, $\tau$, and the parameters of the torus.  This was done in \hms.
The six noncompact scalars from four dimensions lead to a factor of
$\BR^{6r}$.  The two polarizations of the photon lead to two factors of
$\bT(G)$ such that the moduli space is
\eqn\fourtwom{{\BR^{6r} \times \bT(G) \times \bT(G) \over \CW}.}
The metric on the two $\bT(G)$ factors depends on $\tau$ and the
parameters of the compactification.
We can use T duality in the two dimensional theory and convert one or
both of the $\bT(G)$ factors to $\bT({}^*G)$:
\eqn\fourtwomn{\eqalign{
&{\BR^{6r} \times \bT(G) \times \bT({}^*G) \over \CW} \cr
&{\BR^{6r} \times \bT({}^*G) \times \bT({}^*G) \over \CW}.}}
Therefore, S duality,
which exchanges $\tau \rightarrow -{1 \over \tau}$ and $G \rightarrow
{}^*G$ in four dimensions, translates to T duality after compactification
\hms.  Clearly, the physics near the orbifold singularities of
\fourtwom\ or \fourtwomn\ generalizes in an obvious way the discussion of
\dvv. 

\newsec{Theories with $(0,2)$ supersymmetry in $d=6$ and their
compactification} 

Here we study the theories with $(0,2)$ supersymmetry in $d=6$ and their
compactification.  These theories first appeared in the study of K3
compactifications of the Type IIB theory
\ref\zerotwow{E. Witten, hep-th/9507121, ``Some Comments on String
Dynamics,'' Contributed to STRINGS 95: Future Perspectives in String
Theory, Los Angeles, CA, 13-18 Mar. 1995.}
and later in the context of nearby 5-branes in M-theory
\nref\zerotwos{A. Strominger, ``Open P-Branes,'' hep-th/9512059,
\pl{383}{1996}{44}.}%
\nref\wittenbranes{E. Witten, ``Five-branes and M-Theory on an
Orbifold,'' hep-th/9512219, \np{463}{1996}{383}.}%
\refs{\zerotwos,\wittenbranes}.  These theories are expected to be
non-trivial fixed points of the renormalization group in six
dimensions.  Therefore, they have no dimensionful parameter.
Furthermore, since these fixed points are isolated, they have no
dimensionless parameter.

Along the moduli space of the six dimensional theory there are $r$
tensor multiplets of $(0,2)$ supersymmetry.  Each of them includes 5
scalars and a two form $B$, whose field strength $H=dB$ three form is
selfdual.  The one form gauge invariance is subject to some global
identification corresponding to the allowed non-trivial fluxes of $H$.
If there are $r$ fields, the fluxes $\int H_a$ ($a=1,...,r$) through
various three cycles are quantized.  Since $H$ is selfdual, the lattice
of charges of these fluxes is a selfdual lattice.  Therefore, the
Abelian one form gauge invariance along the flat directions is
characterized by a selfdual lattice $\Gamma$.

Interesting order parameters in this theory are the generalizations of
the Wilson loops, which we can call Wilson surfaces.  These are given by
$\exp i \int Q^a B_a$, where the integral is over a two surface and $Q
\in \Gamma$.  Since $H$ is selfdual, these are also the generalizations
of the 'tHooft loop.  The equality between them is possible only when
$\Gamma $ is selfdual.

Some important subtleties associated with the definition of the theory
of such two forms were discussed in 
\ref\wittenmore{E. Witten, ``Five-Brane Effective Action in M-Theory,''
hep-th/9610234}.
Even without supersymmetry or fermions the theory needs a spin structure
for its definition.  Since we are studying the supersymmetric theory we
need a spin structure anyway.  This discussion might interfere with the
conclusion above that $\Gamma$ has to be selfdual\foot{We thank E.
Witten for a useful discussion on this point.}.

The $(0,2)$ supersymmetry constrains the metric on the moduli space to
be locally flat.  The only allowed singularities are orbifold
singularities in the metric.  Hence, the moduli space is 
\eqn\sixdme{{\BR^{5r} \over \CW},}
where $\CW$ is a discrete group.  The theory at the singularities is a
superconformal field theory.  The superconformal algebra includes a
$Spin(5)$ R-symmetry \nahm, which acts on the 5 scalars.

As in section 2, the scalars on the moduli space $\Phi$ have dimension
two.  Their expectation values determine the tension of BPS strings, which
exist in the theory.  The reason for that is that $\Phi$ is in a tensor
multiplet, which includes the two form $B$, and $B$ couples
canonically to strings.  Since the field strength of $B$ is selfdual,
these strings are also selfdual.

Consider the compactification of these theories to five dimensions on
a circle of radius $R_6$.  The kinetic terms for the scalars become
\eqn\scalarkin{\int dx^6 (\partial \Phi)^2 \sim {1 \over R_6} (\partial
\phi)^2} 
where the scalar $\phi=R_6\Phi$ is of dimension one.  This
compactification does not produce more scalars and the moduli space
remains as in \sixdme, $\BR^{5r} \over \CW$.  

These theories flow at long distance to super-Yang-Mills theory.  To
see that, note that the two form $B$ becomes in five dimensions a
vector and a two form.  The self-duality condition in six dimensions
identifies them as dual to each other.  Therefore, every $B$ leads to
one vector field in five dimensions.  As in \scalarkin, the gauge
coupling of the five dimensional theory is
\eqn\fivedga{{1 \over g_5^2} = {1 \over R_6}.}
The scale invariance of the six dimensional theory is explicitly broken
by the scale of the compactification.  This scale determines the
dimensionful gauge coupling of the five dimensional gauge theory
\fivedga. The five dimensional gauge theory is not renormalizable.  It
breaks down at a scale of order ${1 \over g_5^2}$.  Furthermore, as we
discussed in section 2, there is no interacting fixed point of the
renormalization group in five
dimensions.  Therefore, we cannot define the five dimensional gauge
theory as the low energy limit of a five dimensional fixed point.  One
way to define it is to use the six dimensional fixed point
\ref\brs{M. Berkooz, M. Rozali and N. Seiberg, ``Matrix Description of
M-theory on $T^4$ and $T^5$,'' hep-th/9704089, RU-97-23, UTTG-12-97.}
(see also
\ref\rozali{M. Rozali, ``Matrix Theory and U Duality in Seven Dimensions,''
hep-th/9702136.}) 
as we did above.  This definition turns out to be useful in Matrix
theory.  The 
fact that the low energy theory is a gauge theory shows that the lattice
$\Gamma$ and the discrete group $\CW$, which appeared in the data of the
six dimensional theory are the weight lattice and the Weyl group of a
group $G$.

The five dimensional gauge theory has a conserved current
$j={}^*F\wedge F$.
Instantons of the gauge theory are charged BPS particles \seibergdfive.
Their masses are proportional to ${1 \over g_5^2} = {1 \over R_6}$.  The
detailed properties of these instantons depend on the precise way the
theory is defined.  In the context where the five dimensional theory
appears as a compactification of the six dimensional theory, this
relation identifies them as Kaluza-Klein momentum modes around the
compact circle \rozali.

Along the flat directions of the five dimensional theory
there are massive BPS particles (the ``W-bosons'' of the gauge theory),
whose masses are proportional to $\phi$.  These can be interpreted as
the strings of the six dimensional theory wrapping the circle, and hence
their masses are $\phi = R_6\Phi$.  There are also the BPS strings, which
are the the six dimensional strings in the noncompact dimensions.  Their
tensions are $\Phi = \phi/R_6 = \phi/g_5^2$.  This relation identifies
them as being 'tHooft-Polyakov monopole solutions of the gauge theory,
which are strings in five dimensions.

As we said, the six dimensional field theory has string like
excitations.  Can it be formulated as a theory of interacting strings?
The observation above suggests that if this is the case, it is not
simply a string field theory.  In the five dimensional theory the
W-bosons appear as fundamental particles.  The strings are constructed
as classical solutions (magnetic monopoles) in the five dimensional
field theory.  Therefore, they can be interpreted as made out of the
W-bosons.  We should not include in the five dimensional theory both
the W-bosons and the strings as elementary degrees of freedom.
However, from a six dimensional point of view these two excitations
are very similar; they originate from the six dimensional string when
it does or does not wrap the circle.  Therefore, it appears that a
naive string field theory like description of the six dimensional
theory will over-count the elementary degrees of freedom.

When these theories are compactified to four dimensions on a two torus
$\bT_{56}$ with radii $R_{5,6}$ they become $N=4$ theories.  Along the
flat directions we find the $5r$ scalars of the six dimensional theory
and $r$ compact scalars, which arise from the Wilson surface of the
two-forms $B$ on $\bT_{56}$.  These scalars take values on
$\BR^r/\Gamma$, and the scale of this torus is $(R_5R_6)^{-\half}$.  The
moduli space is
\eqn\sixtofo{{\BR^{5r} \times (\BR^r/\Gamma) \over \CW}.}
At the singularities we find an $N=4$ theory labeled by a gauge group
$G$ and the dimensionless coupling constant $\tau$, which is determined
as the complex structure of $\bT_{56}$.  As in five dimensions, $\Gamma$
and $\CW$ are the weight lattice and the Weyl group of $G$.  The
$SL(2,\BZ)$ freedom in the complex structure of this torus translates to
$SL(2,\BZ)$ duality in the field theory \zerotwow.  Since $\Gamma$ is
selfdual, $G={}^*G$.  For such theories the $SL(2,\BZ)$ duality acts
without changing the gauge group.  As in the compactification to
five dimensions, we can identify the W-bosons and the magnetic monopoles
as strings wrapping the two different cycles of $\bT_{56}$ \zerotwow.
They are exchanged by $SL(2,\BZ)$.

We can continue to compactify to three dimensions by adding a circle of
radius $R_4$.  Combining the previous analysis with the discussion of the
compactification from 4 to 3 dimensions above we find the moduli space
$\BR^{5r} \times (\BR^r/\Gamma)\times (\BR^r/\Gamma) \times
(\BR^r/\Gamma) \over \CW$, where the scales of the three tori are
$R_{4,5,6} \over \sqrt{R_4R_5R_6}$ (for simplicity we assume that 
all the angles of the torus are right angles).

The simplest such nontrivial theory is the theory of two 5-branes in
eleven dimensions.  The group $G$ associated with this theory is $U(2)$,
which is selfdual, and $\CW=\BZ_2$.  Before compactification the moduli
space is $\BR^5 \times\BR^5 \over \BZ_2$, where each factors comes from
one 5-brane and the $\BZ_2$ reflects the fact that they are identical.
We now compactify them on a three torus $\bT_{456}$ with radii
$R_{4,5,6}$.  We find the moduli space $(\BR^5 \times \widetilde
\bT_{456})^2 \over \BZ_2$, where the radii of the three torus
$\widetilde \bT_{456}$ are $R_{4,5,6} \over \sqrt{R_4R_5R_6}$.  Note
that it is the same as the moduli space of two 2-branes, which move on
$\widetilde \bT_{456}$.  This is consistent with the duality between
them in eight dimensions
\ref\eightdimd{J.M. Izquierdo, N.D. Lambert, G. Papadopoulos and
P.K. Townsend, ``Dyonic Membranes,'' hep-th/9508177,
\np{460}{1996}{560}; O. Aharony, ``String Theory Dualities From
M-Theory,'' \np{476}{1996}{470}, hep-th/9604103.}. 
Furthermore, the theory at the singularity is exactly that of the
three dimensional $U(2)$ theory.

\bigbreak\bigskip\bigskip

\centerline{\bf Acknowledgments}\nobreak
We thank O. Aharony, T. Banks, M. Berkooz, M. Dine, K. Intriligator,
D. Morrison, S. Shenker, P.  Townsend, C. Vafa and E. Witten for
discussions.  This work was supported in part by DOE grant
DE-FG02-96ER40559.

\listrefs

\end